\documentstyle[12pt,aps,epsfig,prl,preprint]{revtex}
\begin{document}
\raggedbottom
\rm
\title{Branching Ratio Measurement of 
the decay $K_L\rightarrow e^+e^-\mu^+\mu^-$.}
\maketitle
\parindent=0.in
\parskip 0 in
A.~Alavi-Harati$^{12}$,
T.~Alexopoulos$^{12}$,
M.~Arenton$^{11}$,
K.~Arisaka$^2$,
S.~Averitte$^{10}$,
R.F.~Barbosa$^{7,**}$,
A.R.~Barker$^5$,
M.Barrio$^4$,
L.~Bellantoni$^7$,
A.~Bellavance$^9$,
J.~Belz$^{10}$,
R.~Ben-David$^{7}$,
D.R.~Bergman$^{10}$,
E.~Blucher$^4$, 
G.J.~Bock$^7$,
C.~Bown$^4$, 
S.~Bright$^4$,
E.~Cheu$^1$,
S.~Childress$^7$,
R.~Coleman$^7$,
M.D.~Corcoran$^9$,
G.~Corti$^{11}$, 
B.~Cox$^{11}$,
M.B.~Crisler$^7$,
A.R.~Erwin$^{12}$,
R.~Ford$^7$,
A.~Glazov$^4$,
A.~Golossanov$^{11}$,
G.~Graham$^{4}$, 
J.~Graham$^4$,
K.~Hagan$^{11}$,
E.~Halkiadakis$^{10}$,
J.~Hamm$^1$,
K.~Hanagaki$^{8}$,  
S.~Hidaka$^8$,
Y.B.~Hsiung$^7$,
V.~Jejer$^{11}$,
D.A.~Jensen$^7$,
R.~Kessler$^4$,
H.G.E.~Kobrak$^{3}$,
J.~LaDue$^5$,
A.~Lath$^{10,\dagger}$,
A.~Ledovskoy$^{11}$,
P.L.~McBride$^7$,
P.~Mikelsons$^5$,
E.~Monnier$^{4,*}$,
T.~Nakaya$^{7}$,
K.S.~Nelson$^{11}$,
H.~Nguyen$^7$,
V.~O'Dell$^7$, 
M.~Pang$^7$, 
R.~Pordes$^7$,
V.~Prasad$^4$, 
B.~Quinn$^{4,\S}$,
X.R.~Qi$^7$,
E.J.~Ramberg$^7$, 
R.E.~Ray$^7$,
A.~Roodman$^{4}$, 
M.~Sadamoto$^8$, 
S.~Schnetzer$^{10}$,
K.~Senyo$^{8}$, 
P.~Shanahan$^7$,
P.S.~Shawhan$^{4}$,
J.~Shields$^{11}$,
W.~Slater$^2$,
N.~Solomey$^4$,
S.V.~Somalwar$^{10}$, 
R.L.~Stone$^{10}$, 
E.C.~Swallow$^{4,6}$,
S.A.~Taegar$^1$,
R.J.~Tesarek$^{10}$, 
G.B.~Thomson$^{10}$,
P.A.~Toale$^5$,
A.~Tripathi$^2$,
R.~Tschirhart$^7$,
S.E.~Turner$^2$ 
Y.W.~Wah$^4$,
J.~Wang$^1$,
H.B.~White$^7$, 
J.~Whitmore$^7$,
B.~Winstein$^4$, 
R.~Winston$^4$, 
T.~Yamanaka$^8$,
E.D.~Zimmerman$^{4}$
\vspace*{.1 in} 
\footnotesize

$^1$ University of Arizona, Tucson, Arizona 85721 \\
$^2$ University of California at Los Angeles, Los Angeles, California 90095 \\
$^{3}$ University of California at San Diego, La Jolla, California 92093 \\
$^4$ The Enrico Fermi Institute, The University of Chicago, 
Chicago, Illinois 60637 \\
$^5$ University of Colorado, Boulder, Colorado 80309 \\
$^6$ Elmhurst College, Elmhurst, Illinois 60126 \\
$^7$ Fermi National Accelerator Laboratory, Batavia, Illinois 60510 \\
$^8$ Osaka University, Toyonaka, Osaka 560-0043 Japan \\
$^9$ Rice University, Houston, Texas 77005 \\
$^{10}$ Rutgers University, Piscataway, New Jersey 08854 \\
$^{11}$ The Department of Physics and Institute of Nuclear and 
Particle Physics, University of Virginia, 
Charlottesville, Virginia 22901 \\
$^{12}$ University of Wisconsin, Madison, Wisconsin 53706 \\
$^{\dagger}$ To whom correspondence should be addressed. \\
$^{*}$ Permanent address C.P.P. Marseille/C.N.R.S., France \\
$^{**}$Permanent address University of S\~{a}o Paulo, S\~{a}o Paulo, Brazil \\

\vspace*{.1 in}

\centerline{ \bf The KTeV Collaboration}

\normalsize

\draft

\begin{abstract}
We have collected a 43 event sample of the decay
$K_L\rightarrow e^+e^-\mu^+\mu^-$ with 
negligible backgrounds
and measured its branching ratio to be 
$(2.62 \pm 0.40 \pm 0.17)\times 10^{-9} $. We see no evidence
for CP violation in this decay.
In addition, we set the 90\% confidence upper limit on the combined
branching ratios for the lepton flavor violating decays 
$K_L\rightarrow e^{\pm}e^{\pm}\mu^{\mp}\mu^{\mp}$   at
${\mathcal B}(K_L\rightarrow e^{\pm}e^{\pm}\mu^{\mp}\mu^{\mp})
\leq 1.23\times10^{-10}$, assuming a uniform phase space distribution.

\end{abstract}
\vspace{+1.2in}

\parindent=2em
\parskip 0.5em 
We present the  observation of the decay 
$K_L\rightarrow e^+e^-\mu^+\mu^-$ and a measurement of
its branching ratio.
This decay
proceeds entirely through the $K\gamma^*\gamma^*$ vertex
and provides the best
opportunity for its study.
Knowledge of the $K\gamma^*\gamma^*$ vertex is crucial
in order to extract short distance information, including
the CKM matrix element $V_{td}$, from  $K_L\rightarrow\mu^+\mu^-$
decays. We are also able to search for CP violating effects
in $K_L\rightarrow e^+e^-\mu^+\mu^-$, as well as the lepton-flavor
violating decay $K_L\rightarrow e^{\pm}e^{\pm}\mu^{\mp}\mu^{\mp}$.
The previous E799-I experiment 
measured ${\mathcal B}(K_L\rightarrow e^+e^-\mu^+\mu^-$) 
= $(2.9^{+6.7}_{-2.4}) \times 10^{-9}$~\cite{guping} with one event.

There are several predictions for the branching ratio.  Quigg and
Jackson ~\cite{ref:quiggjackson}
have used a Vector Meson Dominance (VMD) model to predict
a value of $2.37 \times 10^{-9}$.  
A phase-space model with both CP conserving  and 
CP violating 
form factors by Uy~\cite{ref:uy} 
predicts values from $(1.63 \pm 0.07) \times 10^{-9}$ 
for a  totally CP conserving decay,
to $(3.67 \pm 0.15) \times 10^{-6}$ for a totally CP violating decay.
Note that the CP violating form factors  increase
the branching ratio by over three orders of magnitude.
This calculation does not take into account any momentum dependence of the
form factors. 
An ${\mathcal O} (p^6)$ Chiral Perturbation Theory 
calculation by Zhang and Goity ~\cite{ref:zhang-goity}
predicts $(1.30 \pm 0.15)\times 10^{-9}$.

The measurement presented here was performed as part of the KTeV 
experiment, which has been described elsewhere~\cite{ref:ktev}. 
The data used were collected during the 1997 run.
The KTeV experiment, as configured for rare decay searches (E799-II), 
used two nearly parallel kaon beams 
created by 800 GeV/c protons incident on a BeO target.  The kaon decays
used in our studies were collected in a decay region approximately
65 m long, situated 94 m from the 
production target.  

Charged particles were detected by four
drift chambers, each consisting of one horizontal and one vertical
pair of planes, with typical resolution of 100 $\mu m$ per plane.
Two drift chambers were situated on either side of an analysis magnet
which  imparted approximately 205 MeV/c of transverse momentum to
the charged tracks.  The drift chambers were followed by a
trigger hodoscope bank and a 1.9 m
$\times$ 1.9 m  calorimeter composed of
3100 blocks of pure CsI.
The fiducial volume was surrounded by a photon veto
system used to reject events in which photons missed the calorimeter.
The calorimeter was followed by a
muon filter composed of 
a 10 cm thick lead wall and three steel walls
totalling 511 cm.  The first plane of scintillators used
to identify muons (MU2) was located after 400 cm of steel,
behind the second steel wall.
Two additional 3 m $\times$ 3 m  scintillator planes
(MU3Y and MU3X), located
after the third steel wall and
consisting of  one horizontal and one
vertical plane,  defined the acceptance for
muons.   
All muon scintillator planes had 15 cm segmentation. 

The trigger 
for the signal events 
required  hits in 
the upstream  drift chambers
consistent with at least two tracks,  as well as two hits in the trigger 
hodoscopes.  The calorimeter was required to have at least one
cluster with over 1 GeV of energy, 
deposited within a  20 ns window relative to the 
event trigger.
The  muon counters MU3X and MU3Y were 
each 
required to have at least two hits. In addition,  
a preliminary online identification of
$K_L\rightarrow e^+e^- \mu^+\mu^-$ decays required a minimum of three
tracks originating from a loosely defined vertex.  
A separate trigger was used to collect 
$K_L\rightarrow \pi^+\pi^-\pi^0$  decays with 
subsequent Dalitz decays $\pi^0\rightarrow e^+e^-\gamma$ 
($K_L\rightarrow \pi^+\pi^-\pi^0_D$) which were used for normalization.  
This trigger was similar
to the signal trigger but had no requirements on 
hits in the muon hodoscopes or clusters in the calorimeter. 
The 
preliminary online identification was performed on the normalization
sample as well.  The normalization mode trigger was
prescaled by a factor of 500:1.

The offline analysis required four tracks from a single vertex.
A cut on the vertex reconstruction $\chi^2$ ensured the four tracks
originated from the same vertex.
Upstream and downstream track segments were allowed at most
a 2 mm offset at the bend plane of the analysis magnet.
The four track decay vertex projected to the calorimeter
was required to be within one of the beam regions.  
A track was identified as $e^{\pm}$ if it pointed to  
a cluster in
the calorimeter with $|E/P-1| \leq 0.05$, otherwise
it was identified as a muon.  
The electromagnetic 
energy resolution of the calorimeter,
$\sigma(E)/E = 0.45\% \oplus 2.0\%/\sqrt{E{\rm ~(GeV)}} $, was  
determined using a large
sample of  $e^{\pm}$ from $K_L\rightarrow\pi^{\mp}e^{\pm}\nu$ decays.

To remove muons which range out in the steel,
events in which a muon track had $P<10$ GeV/c
were discarded.  
To remove misidentified pions, events in which a muon track
deposited over 3 GeV in the calorimeter were discarded.
Events with excessive energy
in the veto counters
beyond that expected from accidental activity, or extra calorimeter
clusters not associated with an electron or muon track,
were also discarded.

\begin{figure}[tbh]	
\centerline{\epsfysize 6.5 truein \epsfbox{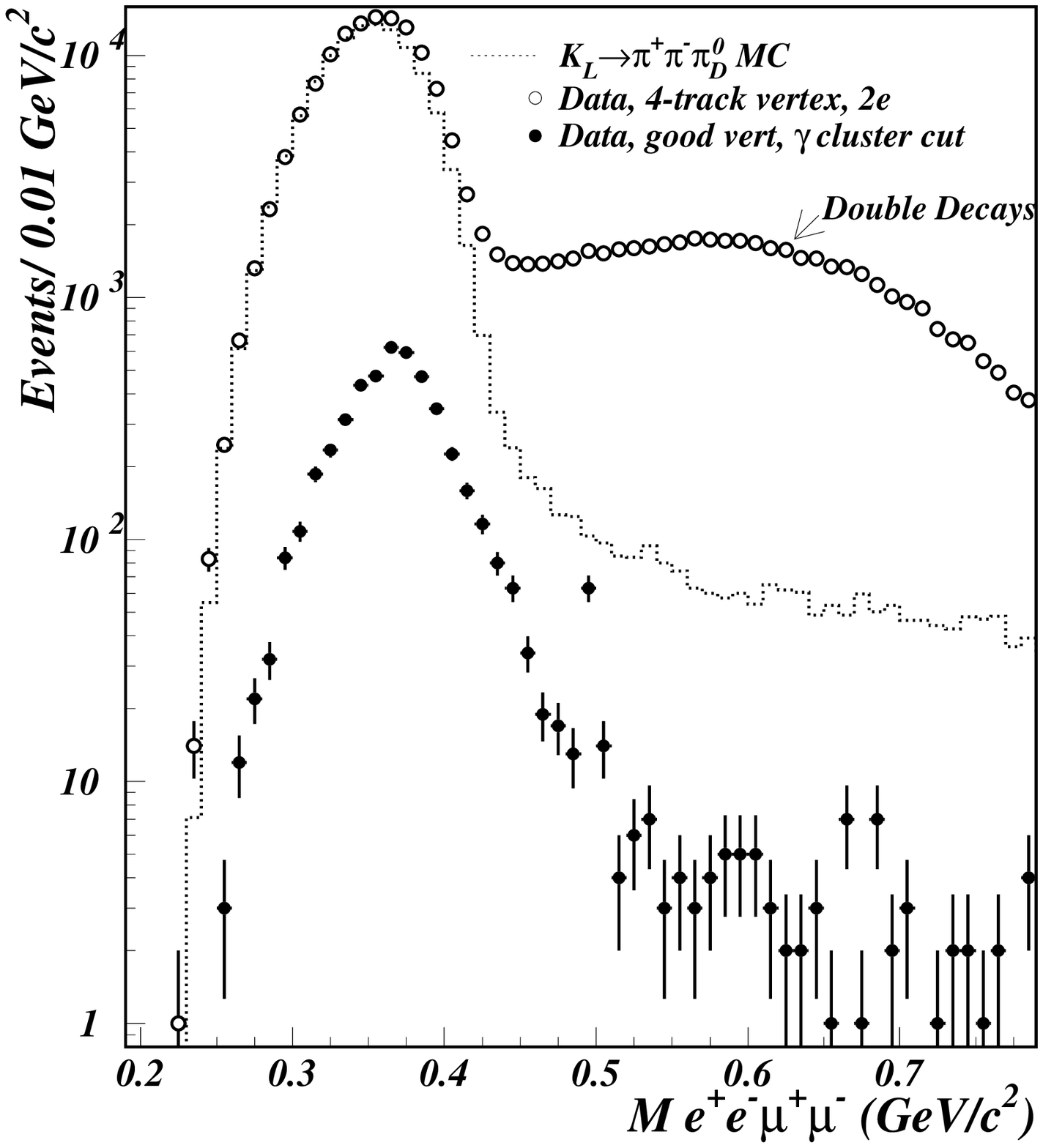}}   
\vskip -0.02 cm
\caption[]{
\label{fig:ovlay}
Distribution of $M_{e e \mu\mu}$ 
after  finding a four-track vertex and
identifying two of the tracks as $e^{\pm}$ 
for data (hollow circles), and for
Monte Carlo simulation of
$K_L\rightarrow \pi^+\pi^-\pi^0_D$
with charged pion punchthrough/decay at the same stage of analysis
(dotted line).
The simulation is normalized to the data below 0.32 GeV/c$^2$.
The data distribution
after cuts on  vertex quality and 
any extra  clusters in the calorimeter is shown 
by filled circles.  A signal peak is visible at the kaon mass.
At this stage of the analysis, $\sim$20\% of the events in that peak 
are $K_L\rightarrow \mu^+\mu^-\gamma$  with $\gamma\rightarrow e^+e^-$ 
in the
vacuum window.}

\end{figure}

A large component of the background consists of
$K_L\rightarrow \pi^+\pi^-\pi^0_D$ decays in which the
charged pions simulate muons by punching 
through the muon filter or decaying in flight.
A simulation of this background
is compared to the data  in figure~\ref{fig:ovlay}.
The invariant mass ($M_{e e \mu\mu}$) distributions for the 
data  and  for the simulated background are shown after successful 
electron identification, but
before cuts on vertex quality or extra calorimeter activity.

The events at high invariant mass are due to
two kaons which decayed within the same  Tevatron
RF bucket  (double decay events).  An event with a double decay contains
two separate two-track vertices, which usually form a poor four-track vertex. 
The vertex quality  cut eliminated nearly
all of the double decay events.  
The  cuts on photon veto activity  and 
extra calorimeter  clusters  removed 98\% of the 
background from $K_L\rightarrow \pi^+\pi^-\pi^0_D$. 
Figure~\ref{fig:ovlay} also shows the data after the vertex
quality and extra calorimeter clusters cut.

Another potentially large background comes from
$K_L\rightarrow \mu^+\mu^-\gamma$ events in which the gamma converts to
an  $e^+e^-$ pair in the vacuum window.
A cut requiring that either the 
two-electron invariant mass ($M_{ee}$) be greater 
than 3 MeV/$c^2$ or  
the separation for the two electron tracks at the first
drift chamber be greater than 3 mm 
reduced the window conversion events 
by 98.8\% while losing  
only 13.6\% of the remaining signal events.    
We estimate  a 0.19 event background
from the vacuum window conversions.  

\begin{figure}[tbh]	
\centerline{\epsfysize 7 truein \epsfbox{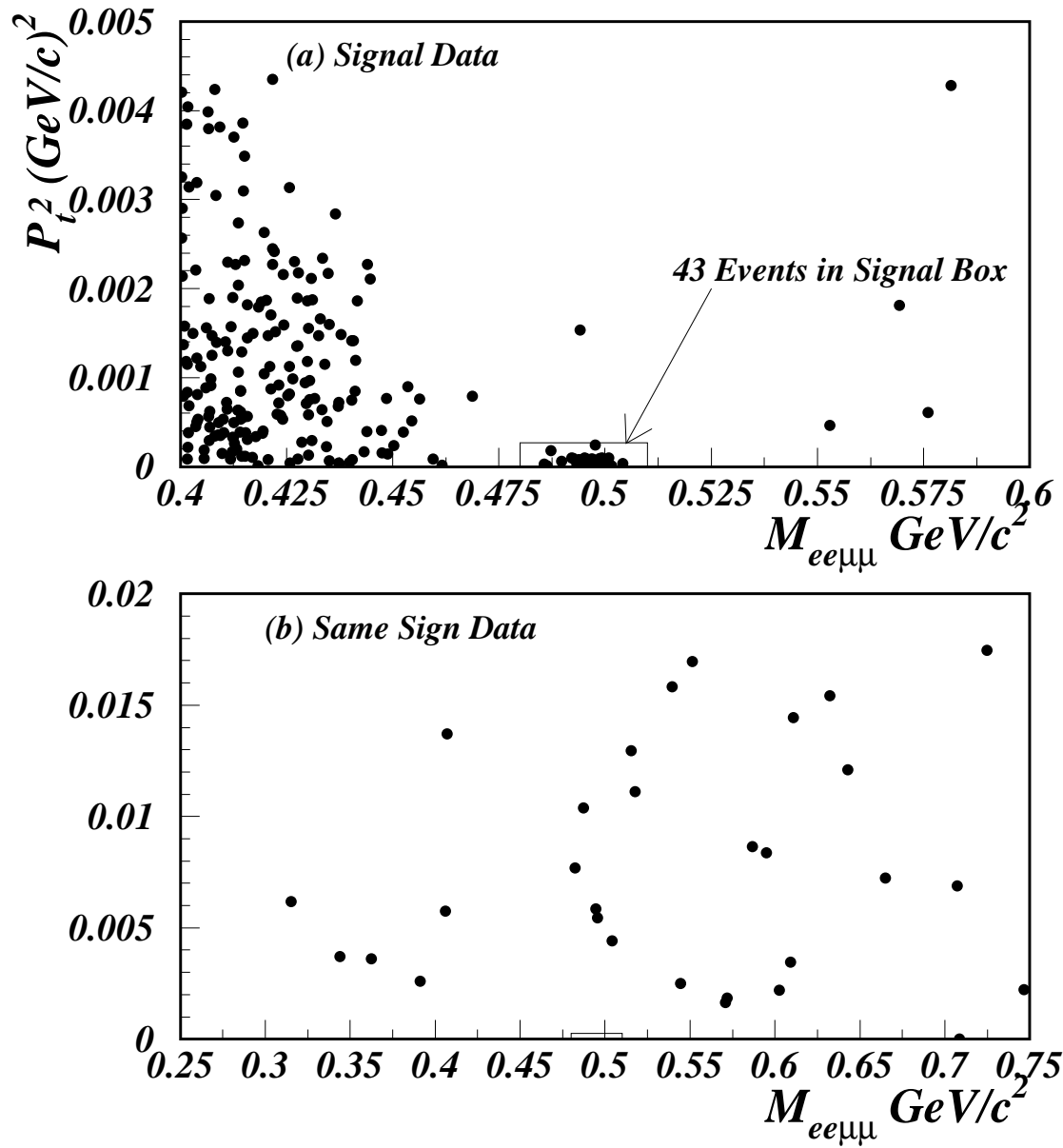}}   
\vskip -0.02 cm
\caption[]{
\label{fig:mvpt2_both}
 Top plot (a) shows 
$P_t^2$ vs. $M_{e e \mu\mu}$ for $K_L\rightarrow e^+e^-\mu^+\mu^-$,
signal events, with all cuts. There are 43 events in the signal box.
Bottom plot (b) shows
same sign lepton events,
$e^{\pm} e^{\pm}$  $\mu^{\mp} \mu^{\mp}$. There are no same sign 
events in the signal box.  Note that the area in plot (b) is a factor
of 10 larger than that in (a).}
\end{figure}

The double decay background remaining after all cuts 
was estimated by examining events in which the  two electrons
were of the same charge sign.  Two electrons (or two muons) 
that do not arise from the same decay  have the
same sign as often as  opposite signs.  
Figure~\ref{fig:mvpt2_both}  shows  scatter plots of 
$M_{e e \mu \mu}$  vs. $P_t^2$, where $P_t^2$ is the square of the 
transverse momentum with respect to the trajectory from the production 
target to the decay vertex. Figure~\ref{fig:mvpt2_both}(a)
 contains signal events with
all but $M_{e e \mu \mu}$ and $P_t^2$ cuts,
while figure~\ref{fig:mvpt2_both}(b) shows the distribution
for events that pass the same cuts, 
except that the two electrons (and muons) are required
to have the same sign.  
There are 30 such like-sign events in the region
$0.25 \leq M_{e  e \mu  \mu}\leq 0.75 {\rm ~GeV/c}^2 $  
and $P_t^2 \leq 0.02 (\rm{~GeV/c})^2$.
We assume that this background is evenly distributed,
and estimate a 0.02 event double decay background 
within the signal region.
The  $M_{e e \mu\mu}$  vs. $P_t^2$ data distribution  shown  in 
figure~\ref{fig:mvpt2_both}(a) 
contains 43  events within
the signal box given by $0.48\leq M_{ee\mu\mu}\leq 0.51 {\rm ~GeV/c}^2$
and $P_t^2\leq 0.00025 {\rm ~(GeV/c)}^2$. 

\begin{figure}

\begin{center}  \hspace*{-0.5in}
   {\epsfysize 7 in\epsffile
                           {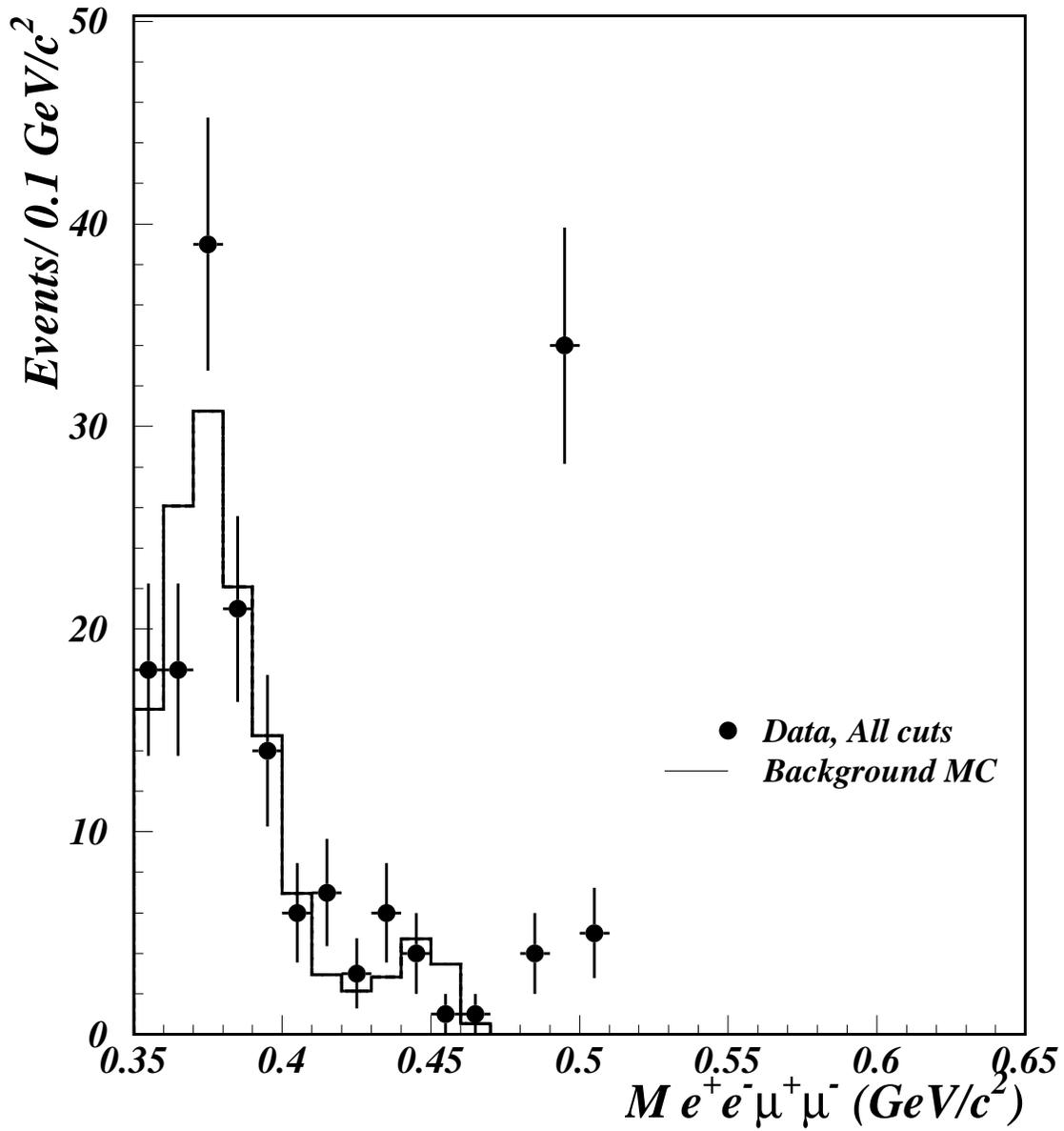}}
\end{center}

\caption{$M_{ee\mu\mu}$ for data (dots) and the 
scaled 
background simulation (line), with $P_t^2\leq 0.00025$ 
(GeV/c)$^2$.  The peak resolution for the signal is approximately 
4 MeV/$c^2$.}
\label{fig:allbckg}
\end{figure}

The remaining background at masses below $M_K$  is mainly from
$K_L\rightarrow \pi^+\pi^-\pi^0_D$ decays
with charged pion decay or punchthrough. 
Figure~\ref{fig:allbckg} shows the fit of the data with all 
but the invariant mass cut to a
scaled background simulation, from which we  
estimate  a background of 0.03 events  in
the signal region.
The total background in the signal region, from window conversions
of $K_L\rightarrow\mu^+\mu^-\gamma$ decays,  
$K_L\rightarrow \pi^+\pi^-\pi^0_D$
decays in which the charged pions  either decay in flight
or punchthrough the filter steel, and  double decay events is 
estimated to be 0.24 events.  

The geometric acceptance of the detector 
for the signal mode, calculated using a Monte
Carlo generator which employed
the matrix element formulated by Uy~\cite{ref:uy}, was  6.1\%. 
Only CP conserving elements of the matrix element were used for the 
acceptance calculation. 

In order to calculate the branching ratio, we use the 
$K_L\rightarrow\pi^+\pi^-\pi^0_D$ decay for normalization.
These events were selected by the minimum-bias, two-track trigger 
described above.  The cuts applied to the
normalization mode were as similar to the signal mode as possible, with a
few separate cuts specific to this mode, 
including the requirement that  the photon
cluster in the calorimeter 
deposit at least 5 GeV of energy in the calorimeter
and be  $\geq 3$ cm away from the beam holes.
We also required that the 
invariant mass of the $e^+ e^-\gamma$  be within 10 MeV/$c^2$ of
the $\pi^0$ mass, and its energy be between 15 and 85 GeV. 
We determined that $(2.70 \pm 0.08) \times 10^{11}$ $K_L$ 
within an energy range of 20 to 220 GeV decayed  between 
90 and 160 meters from the target.  The error in this value
is dominated by the uncertainty in the measured 
branching ratios for $K_L\rightarrow\pi^+\pi^-\pi^0$ and
$\pi^0\rightarrow e^+e^-\gamma$ \cite{ref:PDG}.

While similar in
most respects to the signal trigger, the normalization mode trigger
lacked any muon requirements. 
One source of systematic uncertainty in the normalization 
involved the response of the muon steel and counters.
We evaluated the hit efficiencies of the muon counters 
using  high statistics samples of $\mu^{\pm}$ 
taken in runs with special absorbers and magnet configurations.
The muon efficiency was  $\sim$99\%  and 
nearly uniform across 
the counter planes.  This efficiency was 
measured to $\leq$ 0.5\% of itself.  
We  used similar runs to study our scattering 
simulation. 
By selecting  tracks which traversed an overlap of adjacent 
muon counters in MU2,
we were able to determine the track position at the counter plane
independent of the tracking system, and ensure the
accuracy of the muon scattering simulation for incident muon momenta 
well below the 10 GeV/c cutoff.

We estimated the systematic uncertainty in the branching ratio 
due to any disagreements between the simulation and data by varying 
the selection cuts.
The contribution due to uncertainties in the selection criteria 
for both the signal and normalization
modes is 3\%.   Accidental activity in the
detector caused a 3.7\% uncertainty in the acceptance, while  
systematic uncertainty due to 
muon scattering in the lead and steel downstream of the
calorimeter  was 
estimated to be 0.5\%.  The  change in signal acceptance due to
possible  form-factors in the matrix element of the 
decay contributed 3\%, while the uncertainty due to 
background events remaining in the signal region was estimated at
0.6\%.  

The total systematic uncertainty was 6.4\%, 
to be compared to 15.2\% statistical uncertainty.
Our final result is thus
${\mathcal B} 
(K_L\rightarrow e^+e^-\mu^+ \mu^-) = (2.62 \pm 0.40  
\pm 0.17) \times 10^{-9}$, where the first error is statistical and the
second is systematic.

Events with like-sign leptons are a signature of lepton flavor
violation, and given that we found no like-signed lepton events
in our sample, we  put a limit on the process 
$K_L\rightarrow e^{\pm}e^{\pm}\mu^{\mp}\mu^{\mp}$.   Using a 
flat phase-space generator, similar analysis requirements
and signal region  as that
used for the $e^+e^-\mu^+\mu^-$ analysis, 
we found that the geometric acceptance for
like-sign lepton events was 7.3\%. With zero events in the 
signal region,  we find that the combined 
${\mathcal B}(K_L\rightarrow e^{\pm}e^{\pm}\mu^{\mp}\mu^{\mp})\leq 
1.23\times10^{-10}$
at the 90\% confidence level.

Uy~\cite{ref:uy} has carried out
a phase space calculation for $K_L\rightarrow e^+e^- \mu^+\mu^-$, 
including  
a CP violating term in the decay.  The form factors used
had no $q^2$ dependence and were labelled $g_2$ for the CP even and $h_2$
for the CP odd component.
According to this model, the total measured branching ratio is 
sensitive to the CP violating term.
We measure the ratio $(g_2/h_2)^2 \leq 2.71\times 10^{-4}$ with 90\% 
confidence.

The angular distributions of the decay products are also a 
sensitive probe of CP violating effects.  Define the angle
$\phi$ to be the angle between the plane containing the 
$e^+e^-$ and that containing the $\mu^+\mu^-$.  An asymmetry
about zero in the distribution $\sin2\phi$ is unambiguous evidence
for CP violation.  We have observed such an 
asymmetry in the
decay $K_L\rightarrow\pi^+\pi^-e^+e^-$~\cite{ref:ppee}.
For the decay  $K_L\rightarrow e^+e^-\mu^+\mu^-$,
18 (25) events 
are in the positive (negative) 
$\sin2\phi$ direction, consistent with zero asymmetry.

\begin{figure}

\begin{center} \hspace*{-0.5in}
   {\epsfysize 7 in\epsffile
                           {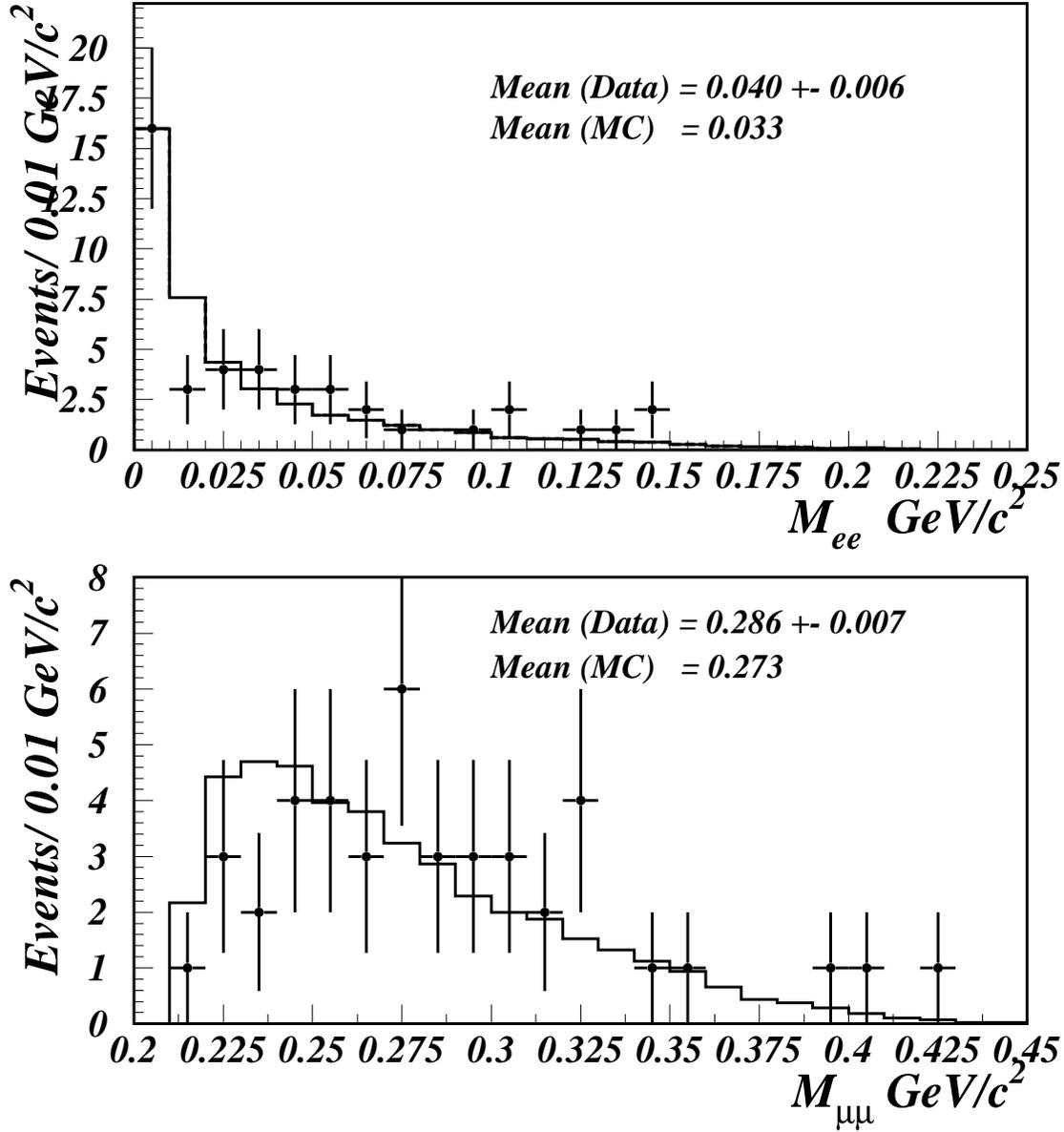}}
\end{center}
\vskip -0.02 cm

\caption{The distributions of $M_{ee}$ and $M_{\mu\mu}$ for the 43 events
observed in the data (dots) as well as that expected from a model
without any momentum dependent form factors (line).}
\label{fig:formfact}
\end{figure}

In principle, the $K_L\rightarrow e^+e^-\mu^+\mu^-$ decay is the best 
mode to investigate the possible $K\gamma^*\gamma^*$ form factors because
there are no exchange terms to complicate the theoretical understanding.
However, the low statistics makes determination of the form factors
difficult at present.  
Figure~\ref{fig:formfact}
shows the distributions for $M_{ee}$ and  $M_{\mu\mu}$ for the 43 signal
events, along with the shape expected from a simulation without any
momentum dependence in the form factors.  
Although there may be a discrepancy between data and MC prediction
in the $M_{\mu\mu}$ distribution indicative of the presence of a form factor, 
the lack of statistics makes a firm conclusion difficult at this time.

We have measured the branching ratio for the decay 
$K_L\rightarrow e^+e^-\mu^+ \mu^-$ to be 
$(2.62 \pm 0.40{\rm ~(stat)} \pm 0.17{\rm ~(sys)}~)\times 10^{-9}$.
Note that the VMD model of Quigg and Jackson~\cite{ref:quiggjackson} 
which predicts $2.37 \times 10^{-9}$
is in better agreement with our result than more recent predictions.
We have placed an upper limit on the CP violating term 
of Uy~\cite{ref:uy}, of $(\frac{g_2}{h_2})^2\leq 2.71\times 10^{-4}$
at the 90\% C.L, using the branching ratio.
We have also searched for  lepton flavor violation in
like-sign lepton decays,  and placed a limit
${\mathcal B}(K_L\rightarrow e^{\pm}e^{\pm}\mu^{\mp}\mu^{\mp})\leq 
1.23\times10^{-10}$
at the 90\% confidence level.
The 1999 run of KTeV will soon yield a larger statistics sample
of $K_L\rightarrow e^+e^-\mu^+ \mu^-$ decays.  Studies of form
factors in these decays should advance the understanding of the
$K\gamma^*\gamma^*$ vertex, thereby constraining one of the
largest sources of uncertainty in the determination of the
CKM matrix element $V_{td}$ from $K_L\rightarrow \mu\mu$ decays.
In addition, a CP violating contribution might be detected with 
a larger sample of  $K_L\rightarrow e^+e^-\mu^+ \mu^-$ decays.

We gratefully acknowledge the support and effort of the Fermilab
staff and the technical staffs of the participating institutions for
their vital contributions.  This work was supported in part by the U.S. 
Department of Energy, the National Science Foundation and the Ministry of
Education and Science of Japan.


\end{document}